\documentclass[conf]{IEEEtran}
\usepackage{algorithm,algpseudocode}
\usepackage{amsmath}
\usepackage{amssymb}
\usepackage{mathtools}

\usepackage{cite}
\def\BibTeX{{\rm B\kern-.05em{\sc i\kern-.025em b}\kern-.08em
    T\kern-.1667em\lower.7ex\hbox{E}\kern-.125emX}}
\usepackage[bookmarksnumbered=true]{hyperref} 
\hypersetup{
     colorlinks = true,
     linkcolor = blue,
     anchorcolor = blue,
     citecolor = blue,
     filecolor = blue,
     urlcolor = blue
     }

\usepackage{svg}

\usepackage{wrapfig}
\usepackage{caption}
\usepackage{subcaption}
\usepackage{booktabs}

\usepackage{tikz}
\def\checkmark{\tikz\fill[scale=0.5](0,.35) -- (.25,0) -- (1,.7) -- (.25,.15) -- cycle;} 

\newcommand{\Lagr}{\mathcal{L}}
\usepackage{xcolor}

\begin{document}

\title{HyDe: A \underline{Hy}brid PCM/FeFET/SRAM \underline{De}vice-search for Optimizing Area and Energy-efficiencies in Analog IMC Platforms}

\author{
Abhiroop Bhattacharjee, \textit{Student Member, IEEE}, Abhishek Moitra, \textit{Student Member, IEEE}, and Priyadarshini Panda, \textit{Member, IEEE} \\
\{abhiroop.bhattacharjee, abhishek.moitra, priya.panda\}@yale.edu \\

Department of Electrical Engineering, Yale University, USA

}

\maketitle

\begin{abstract}

Today, there are a plethora of In-Memory Computing (IMC) devices- SRAMs, PCMs \& FeFETs, that emulate convolutions on crossbar-arrays with high throughput. Each IMC device offers its own pros \& cons during inference of Deep Neural Networks (DNNs) on crossbars in terms of area overhead, programming energy and non-idealities. A design-space exploration is, therefore, imperative to derive a hybrid-device architecture optimized for accurate DNN inference under the impact of non-idealities from multiple devices, while maintaining competitive area \& energy-efficiencies. We propose a two-phase search framework (HyDe) that exploits the best of all worlds offered by multiple devices to determine an optimal hybrid-device architecture for a given DNN topology. Our hybrid models achieve upto $2.30-2.74\times$ higher $TOPS/mm^2$  at $22-26\%$ higher energy-efficiencies than baseline homogeneous models for a VGG16 DNN topology. We further propose a feasible implementation of the HyDe-derived hybrid-device architectures in the 2.5D design space using chiplets to reduce design effort and cost in the hardware fabrication involving multiple technology processes.

\end{abstract}

\begin{IEEEkeywords}
In-Memory Computing, Hybrid-device Architecture, Non-idealities, Area \& Energy-efficiency, 2.5D chiplet-integration
\vspace{-4mm}
\end{IEEEkeywords}

\IEEEpeerreviewmaketitle

\section{Introduction}
\label{sec:intro}

Deep Neural Networks (DNNs) have become ubiquitous for a wide range of applications ranging from computer vision, voice recognition to natural language processing \cite{alzubaidi2021review}. Today, In-Memory Computing (IMC) has become popular as an alternative platform addressing the ‘memory-wall’ bottleneck of von-Neumann architectures for energy-efficient implementation of DNNs \cite{sebastian2020memory}. Specifically, analog crossbars based on a plethora of emerging (beyond-SRAM) IMC devices such as RRAMs, PCMs, FeFETs have been widely researched for compact, energy-efficient and accurate inference of DNNs \cite{chakraborty2020pathways}.

\begin{figure*}[t]
    \centering
    \includegraphics[width=\linewidth]{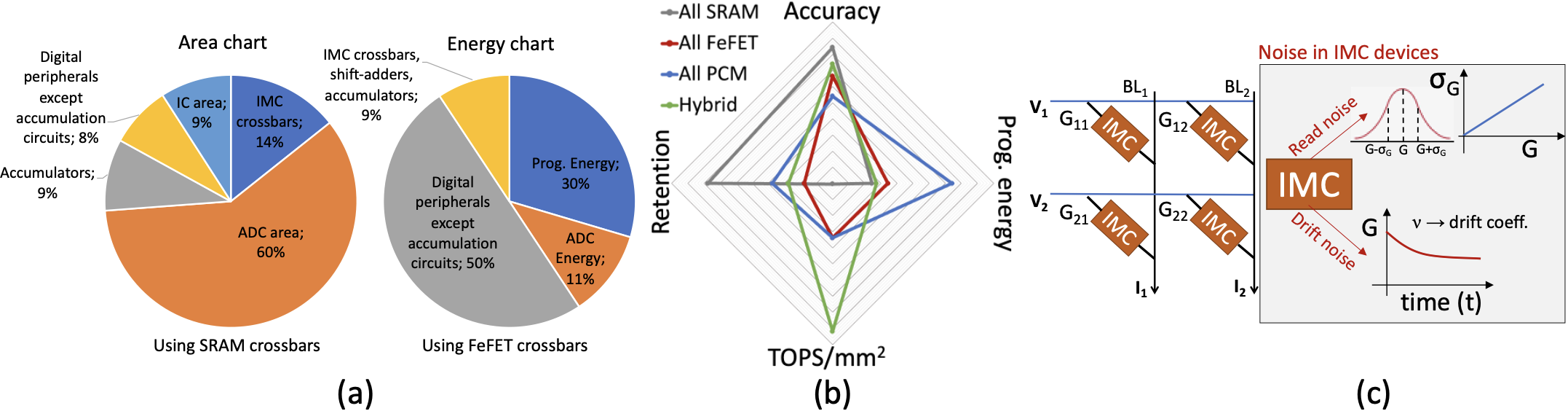}
   
    \caption{(a) Pie-charts plotted using Neurosim tool \cite{chen2018neurosim} showing on-chip area \& energy distributions when a VGG16 DNN is deployed on IMC crossbars of size 128$\times$128. (b) Radar-chart illustrating the trade-offs existing in homogeneous device-based IMC architectures and how a HyDe-derived hybrid architecture can attain optimal performance in terms of accuracy, area-efficiency, programming cost and retention. Note, this chart is for representation purpose and is not to scale. (c) A 2$\times$2 IMC crossbar with the IMC device-level noises highlighted.}
    \label{intro_fig}

\end{figure*}

Owing to device non-idealities and limited device precisions greatly impacting analog-based computing and the high cost of programming associated with the emerging IMC devices, traditional SRAM-based IMCs are used for accurate DNN inference on hardware \cite{spetalnick2022practical, krishnan2022hybrid}. However, as shown in Fig. \ref{intro_fig}(a) (Area chart) for a VGG16 DNN architecture mapped onto an SRAM-based IMC hardware, the SRAM crossbars account for a significant portion of the overall chip area ($\sim14\%$) \cite{chen2018neurosim}. Now, suppose the same DNN is mapped onto an IMC hardware based on FeFET crossbars, then the area expenditure of the IMC crossbars is $<1\%$ of the overall chip area. However, this compactness comes at the cost of increased device-level non-idealities as well as higher device-programming costs \cite{sebastian2020memory}. Furthermore, emerging devices like FeFETs have poor retention capabilities ($\sim10^4s$), and thus need to be frequently re-programmed to overcome accuracy degradation on hardware \cite{byun2022recent}. Thus, the device-programming energy becomes an important component of the total inference energy on hardware. As shown in Fig. \ref{intro_fig}(a) (Energy chart) for the VGG16 DNN architecture mapped onto an FeFET-based IMC hardware, programming energy accounts for $\sim30\%$ of the total chip-energy. PCM devices, albeit have better retention, are more susceptible to high device-to-device variations during read operations \cite{nandakumar2018phase, byun2022recent} resulting in reduced inference accuracies on hardware. Clearly, we see an accuracy-energy-area trade-off where, each IMC device has its own pros and cons and there is no clear winner when a full DNN model is deployed on IMC hardware using a single device (see Fig. \ref{intro_fig}(b)). Thus, device-level heterogeneity in an IMC architecture can exploit the best of all the worlds. This can help achieve high area-efficiency \& inference accuracy resilient to the impact of device-specific non-idealities, while maintaining reasonably high retention at lower programming costs. 

Recently there have been works based on hybrid RRAM/SRAM IMC architectures. \cite{rashed2021hybrid, luo2020accelerating} improve the noise-resilience of IMC hardware during training by decomposing dot-product computations into digital boolean operations and analog Vector-Matrix-Multiplications (VMMs). This is done using SRAMs to encode the MSBs and RRAMs for the LSBs of multi-bit weights, while maintaining reasonably high energy-efficiency. \cite{you2020new} is also based on hybrid synapses (using FeFET and SRAM), wherein the fundamental structure of an SRAM cell is altered by embedding the FeFET device inside an SRAM cell. This is unlike other types of hybrid synapse works such as \cite{luo2020accelerating, rashed2021hybrid} that require an ensemble of SRAM and an emerging device for encoding multi-precision weights. \cite{krishnan2022hybrid, behnam2022algorithm}, on the other hand,  propose a heterogeneous IMC architecture for training DNNs that integrates an analog RRAM-based crossbar macro with a small digital SRAM macro to compensate for the RRAM-level noises and recover the performance accuracy, with minimal hardware overhead. A recent work called ReHy \cite{jin2021rehy} proposes a hybrid-device neural training accelerator, which performs forward propagation using RRAM-based IMCs, and computes gradients during backward propagation with digital SRAM macros at floating point precision.

All the above works based on hybrid-device IMC are essentially optimizing the hardware in isolation for better noise-resilience and energy-efficiency during training or inference. However, these works are agnostic to the topology of the DNN model to be mapped onto the IMC hardware. The layer configurations of a DNN as well as the trained weights contained in these layers impact the DNN noise-resiliency, area and energy expenditures on IMC-based hardware differently. This motivates us to introduce device-level heterogeneity into our hardware architecture for DNNs, wherein each convolutional layer has affinity for one specific type of IMC device. 

Our methodology of searching the design space to derive optimal hybrid-device IMC architectures (based on PCM/FeFET/SRAM devices) for a given DNN topology is termed as HyDe. We choose PCM, FeFET \& SRAM-based IMC devices as they vary greatly in terms of IMC area, ON/OFF ratios, noise characteristics, retention capabilities, programming costs, thereby providing HyDe with the scope for optimization across multiple conflicting objectives (see Fig. \ref{intro_fig}(b)). While there has been a body of works based on Neural Architecture Search (NAS) \cite{cai2019once, cai2018proxylessnas, jiang2020hardware, choi2021dance, lyu2023designing, bhattacharjee2023xplorenas, moitra2023xpert} that optimize DNN topologies to meet the constraints of a given underlying hardware architecture (CPU, GPU, digital systolic-arrays, IMC and so forth) without incurring accuracy losses, this work for the first time delves deeper into the IMC device-level properties to search for an optimal combination of IMC devices to efficiently realize a fixed DNN topology on hardware. We also propose the method to implement our HyDe-derived hybrid-device models on a IMC crossbar-based monolithic chip for feasible heterogeneous integration. Furthermore, with the advent of the 2.5D/3D roadmap to solve the existing bottlenecks of heterogenenous integration in terms of cost and scalability by using chiplet-technology \cite{krishnan2021siam, murali2020heterogeneous, loh2021understanding}, our proposed vision to implement DNNs on a multi-device IMC architecture can become highly pragmatic.

In summary, the key contributions of this work are:

\begin{itemize}
    \item We propose HyDe framework for searching layer-specific hybrid-device IMC architectures (based on PCM, FeFET and SRAM devices) for noise-tolerant (accurate), area- \& energy-efficient implementations of a fixed DNN model. The hybrid-device architectures have optimal retention capabilities at reasonably low programming costs (see Fig. \ref{intro_fig}(b)). Please note that the HyDe framework is not limited to searching hybrid-device IMC architectures based on PCM, FeFET and SRAM devices only but can be used for any other IMC device whose properties (such as, precision, ON/OFF ratio, noise-profile, programming energy, \textit{etc.})  are input to the framework. 
    
    \item We also find that a layerwise hybrid-device configuration also translates to layer-specific hybrid ADC precision, leading to increased $TOPS/mm^2$ and energy-efficiency during inference. 
    
    \item We use a VGG16 DNN model for CIFAR10 \cite{cifar} \& TinyImagenet \cite{le2015tiny} tasks/datasets for our experiments. For the TinyImagenet dataset, we find that HyDe-searched hybrid model can achieve close to `All SRAM' inference performance accuracy of 52.62\% and $TOPS/mm^2$ as high as 6.49, with optimally high retention ($>10^6s$) at a reasonably low device-programming cost ($110.1\mu J$).  
    
    \item We also find that the HyDe-derived hybrid-device configurations searched using CIFAR10 dataset transfer well to the more complex TinyImagenet dataset, thereby alleviating the need for chip re-fabrication for inferring a different task of the same kind.

    \item We finally propose a pragmatic implementation of the HyDe-derived hybrid-device architectures in the 2.5D design space using an ensemble of device-specific IMC chiplets. 
    
\end{itemize}

\section{Background}

\subsection{IMC Crossbar Architectures}
\label{sec:back_xbar}

Analog crossbars consist of 2D arrays of IMC devices, Digital-to-Analog Converters (DACs), Analog-to-Digital Converters (ADCs) and a programming circuit. The synaptic devices at the cross-points are programmed to a particular value of conductance (between $G_{MIN}$ and $G_{MAX}$) during inference. For dot-product operations, the DNN activations are fed in as analog voltages $V_i$ to each row using DACs and weights are programmed as synaptic device conductances ($G_{ij}$) at the cross-points as shown in Fig. \ref{intro_fig}(c) \cite{jain2020rxnn, bhattacharjee2021neat, sebastian2020memory, buchel2022gradient}. For an ideal crossbar array, during inference, the voltages interact with the device conductances and produce a current (governed by Ohm's Law). Consequently, by Kirchoff's current law, the net output current sensed by ADCs at each column $j$ or Bit-line (BL) is the sum of currents through each device, \textit{i.e.} $I_{j(ideal)} = \Sigma_{i}^{}{G_{ij} * V_i}$.

\textbf{Impact of non-idealities:} In reality, the analog nature of the computation leads to various hardware noise or non-idealities, such as interconnect parasitic resistances, IMC device-level variations and so forth\cite{jain2020rxnn, sun2019impact}. 
Consequently, the net output current sensed at each column $j$ in a non-ideal scenario becomes $I_{j(non-ideal)} = \Sigma_{i}^{}{G_{ij}' * V_i}$, which deviates from its ideal value. This manifests as accuracy degradation for DNNs mapped onto crossbars. In this work, we deal with two specific device non-idealities during inference- read noise and temporal drift noise (illustrated in Fig. \ref{intro_fig}(c)). IMC devices exhibit device-to-device variations in their conductances $G$ that constitute the read noise $\tilde{n} \propto \mathcal{N}(0,\sigma^{2})$, where $\sigma$ signifies the standard-deviation of noise \cite{sun2019impact}. The noisy conductance $G'$ can be written as:

\small
\begin{equation} G'=G+\tilde{n}.
\label{eq:read-noise}
\end{equation}
\normalsize

Memristor devices, like PCMs and FeFETs, are also susceptible to temporal drift in their conductances and with the passage of time, their programmed conductance decreases and approaches higher resistance states (see Fig. \ref{intro_fig}(c)) \cite{sun2019impact}. The effect of temporal drift can be modelled using the equation:

\small
\begin{equation} G'=G*(\frac{t}{t_0})^{-\nu}.
\label{eq:drift-noise}
\end{equation}
\normalsize

Here, $t$ denotes time elapsed since programming the device to conductance $G$ at time $t_0$ ($t_0$ is assumed to be 1s) and $\nu$ denotes the device-specific drift coefficient. Higher the value of $\nu$, poorer the retention of the device during inference. Note, in our experiments $t$ denotes the inference time at which accuracy of a DNN is measured and $t = 0$ denotes the time when initial inference accuracy is measured without taking the impact of temporal drift noise. 

\section{Methodology and Implementation}

\subsection{Parent Architecture}

\label{sec:parent_arch}

HyDe searches for a layerwise optimal hybrid device configuration for a given DNN and task from a parent architecture. As shown in Fig. \ref{overall_fig}(a), the parent architecture is constructed from the fixed DNN model (VGG16 in this work), wherein each 2D convolution (conv2D) operation is replaced with a \textit{Composite Device-aware Convolution} (CompDC) operation. If we use $n$ different kinds of devices, then the output of a CompDC operation is a weighted-sum of the outputs of device-aware conv2D operations using each of the $n$ devices. Each  constituent device-aware convolution in a CompDC layer is associated with a \textit{affinity-parameter} ($\alpha$). Let us assume that a CompDC layer is associated with the affinity-parameters $\alpha_1$, $\alpha_2$ and $\alpha_3$ corresponding to three devices- SRAM, PCM and FeFET. If $p_1$, $p_2$ and $p_3$ are respectively the \textit{softmax} of $\alpha_1$, $\alpha_2$ and $\alpha_3$ and $o_1$, $o_2$ and $o_3$ are the outputs of the device-aware convolutions for the individual devices, then the output of the CompDC operation ($m_{CompDC}$) is computed as:

\small
\begin{equation} m_{CompDC} = \sum_{j=1}^{3}p_j * o_j.
\label{eq:out-CompDC}
\end{equation}

\normalsize

where, 

\small
\begin{equation} p_j = \frac{exp(\alpha_j)}{\sum_{j=1}^{3} exp(\alpha_j)}.
\label{eq:softmax}
\end{equation}

\normalsize

Here, we denote $p_j$ $\epsilon$ [0,1] corresponding to each device constituting the CompDC operation as \textit{probability coefficient}. Our objective is to train the affinity-parameters  corresponding to every CompDC layer in the parent architecture (see Section \ref{sec:hyde_train}) and finally, sample the device with the highest affinity for every DNN layer, and, thereby obtain a hybrid-device configuration. 

\subsection{System Implementation}
\label{sec:sys_imp}

\begin{figure}[t]
    \centering
    \includegraphics[width=\linewidth]{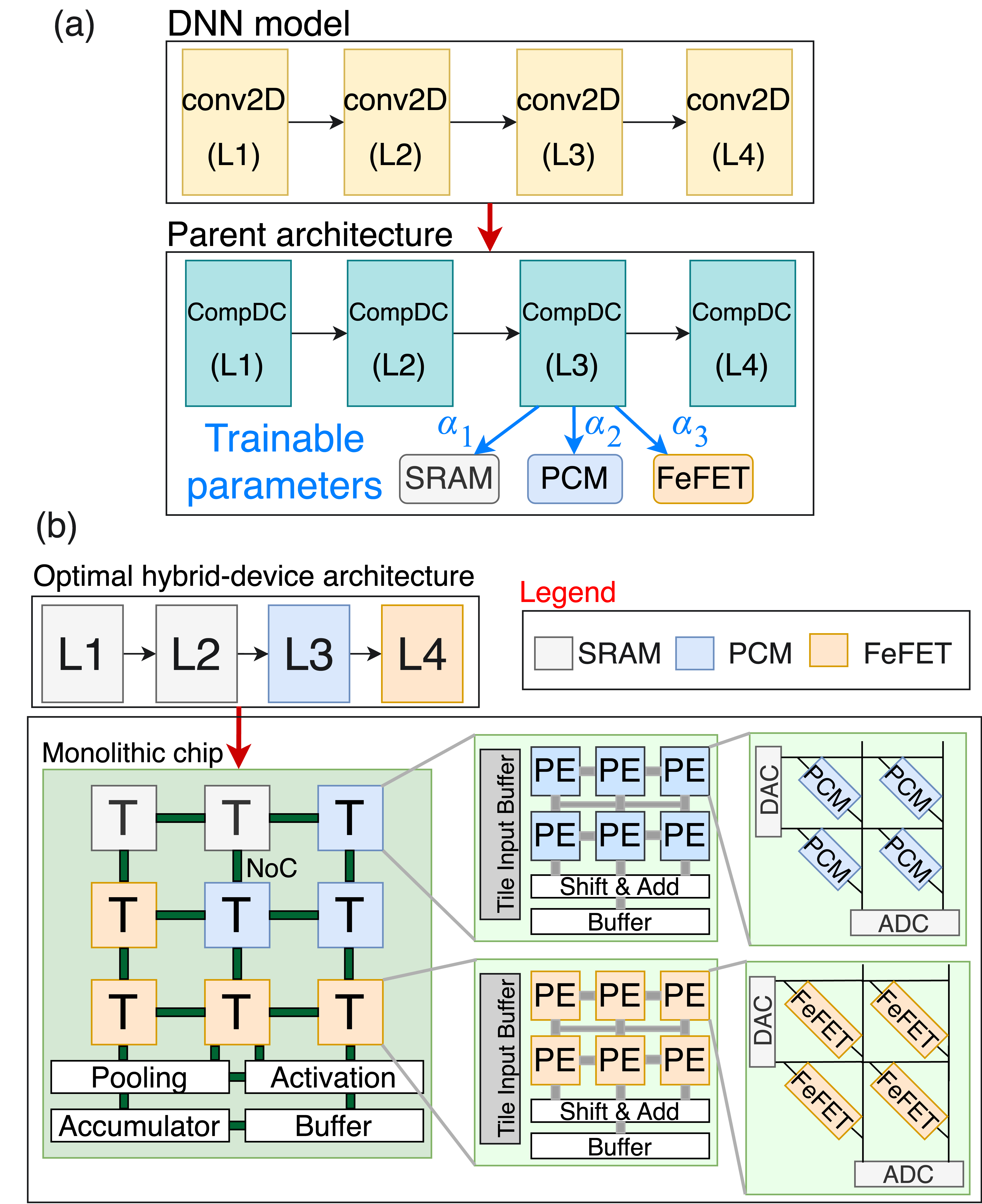}
   
    \caption{(a) Representation of a 4-layered DNN model and its corresponding parent architecture, obtained by replacing conv2D layers with CompDC layers. (b) Mapping of HyDe-derived hybrid-device architecture on a monolithic chip for inference. Grey layers map onto tiles having SRAM PEs, blue layers onto tiles with PCM PEs and orange layers onto tiles with FeFET PEs.}
    \label{overall_fig}
     
\end{figure}

\begin{table*}[t]
\centering
\caption{Device properties used for our HyDe-based hybrid IMC device configuration search. $G$ denotes the programmed device conductance in $\mu S$. N.A. stands for Not Applicable.}
\label{tab:dev_prop}
\resizebox{.75\linewidth}{!}{%

\begin{tabular}{|c|c|c|c|c|c|c|}
\hline
\textbf{Device} &
  \textbf{\begin{tabular}[c]{@{}c@{}}$R_{ON}$, \\ ON/OFF \\ ratio\end{tabular}} &
  \textbf{Precision} &
  \textbf{\begin{tabular}[c]{@{}c@{}}Read \\ noise \\ ($\sigma$)\end{tabular}} &
  \textbf{\begin{tabular}[c]{@{}c@{}}Drift \\ coeff. \\ ($\nu$)\end{tabular}} &
  \textbf{\begin{tabular}[c]{@{}c@{}}Prog. \\ Energy \\ per level\end{tabular}} &
  \textbf{\begin{tabular}[c]{@{}c@{}}Feature \\ area \\ ($F^2$)\end{tabular}} \\ \hline
\begin{tabular}[c]{@{}c@{}}PCM\\ \cite{nandakumar2018phase, haensch2022co}\end{tabular}   & \begin{tabular}[c]{@{}c@{}}40 $k\Omega$,\\ 40\end{tabular}    & 4 bits & $\sigma = 0.03G + 0.13$ & 0.04 & $\sim10 pJ$ & 4           \\ \cline{1-2} \cline{4-7} 
\begin{tabular}[c]{@{}c@{}}FeFET\\ \cite{wang2020ferroelectric, haensch2022co, byun2022recent}\end{tabular} & \begin{tabular}[c]{@{}c@{}}222.22 $k\Omega$, \\ 100\end{tabular} &                         & $\sigma =0.1$           & 0.1  & $\sim2 pJ$  & 6           \\ \cline{1-2} \cline{4-7} 
\begin{tabular}[c]{@{}c@{}}SRAM\\ (1 bit/cell) \cite{jaiswal20198t}\end{tabular}  & \begin{tabular}[c]{@{}c@{}}5 $k\Omega$,\\ $\infty$ \end{tabular}                                                   &                         & $\sigma =0.05$          & N.A. & $\sim10 fJ$ & 120*4 = 480 \\ \hline
\end{tabular}%
}
\end{table*}

Before delving into the training methodology of HyDe, we first explain how to map a hybrid-device configuration onto a hierarchical analog crossbar-based monolithic chip \cite{chen2018neurosim} for inference (see illustration in Fig. \ref{overall_fig}(b)). The entire chip consists of an array of Tiles (T) that are inter-connected using a Network-on-Chip (NoC), along with buffers, accumulators, activation units and pooling units implemented digitally. Within each tile, we have an array of IMC Processing Engines (PEs) inter-connected using a H-Tree and other digital peripherals. Each PE, consisting of one analog crossbar-array of a fixed size and its peripheral circuits such as DACs and ADCs, generates dot-product outputs. Note, each tile consists of a fixed number of PEs (and hence, crossbars) and the topology of a given DNN model dictates the total number of tiles required by the chip to map the entire neural network. Within a tile, all the crossbar-arrays (existing inside the PEs) have synapses of the same device-type (\textit{i.e.}, SRAM, PCM or FeFET). No two layers of a DNN can be mapped onto the PEs inside a single tile \cite{chen2018neurosim}. 

Let us suppose that our hybrid-device configuration for a 4-layered DNN model (see Fig. \ref{overall_fig}) is such that the first two-layers (L1 and L2) have affinity towards SRAM crossbars, while L3 and L4 layers have affinity towards PCM and FeFET crossbars, respectively. Based on the DNN topology, let us consider that L1 and L2 need 1 tile each to be mapped, while L3 and L4 need 3 and 4 tiles, respectively. Thus, the monolithic chip would consist of 2 SRAM-based tiles (grey), 3 PCM-based tiles (blue) and 4 FeFET-based tiles (orange) as shown in Fig. \ref{overall_fig}(b). Note, the digital peripheral circuits around the array of PEs inside a tile and around the array of tiles in the chip are not disturbed by the device-level heterogeneity across tiles and hence, need not be custom designed.

For our HyDe search, the properties of the different IMC devices (SRAM, PCM \& FeFET) used for hybrid-mapping on crossbars are listed in Table \ref{tab:dev_prop}. SRAM crossbars are less impacted by noise, have the highest retention and least programming energy, making them prospective candidates for accurate inference. However, 
SRAMs consume $\sim100\times$ higher IMC area in comparison to other memristor devices. PCMs and FeFETs enable compactness during inference by reducing IMC area but are impacted by read noise and higher programming energies. Specifically, FeFETs suffer from the problem of lower retention ($\sim10^4 s$) due to higher temporal drift coefficient for which they need to be periodically re-programmed for accurate inference \cite{byun2022recent}. PCMs, on the other hand, have better retention capabilities ($>10~years$) \cite{byun2022recent}. However, PCM is more prone to the impact of read noise (due to larger $\sigma$ and smaller ON/OFF ratio) than FeFET.

\begin{figure*}[t]
    \centering
    \includegraphics[width=\linewidth]{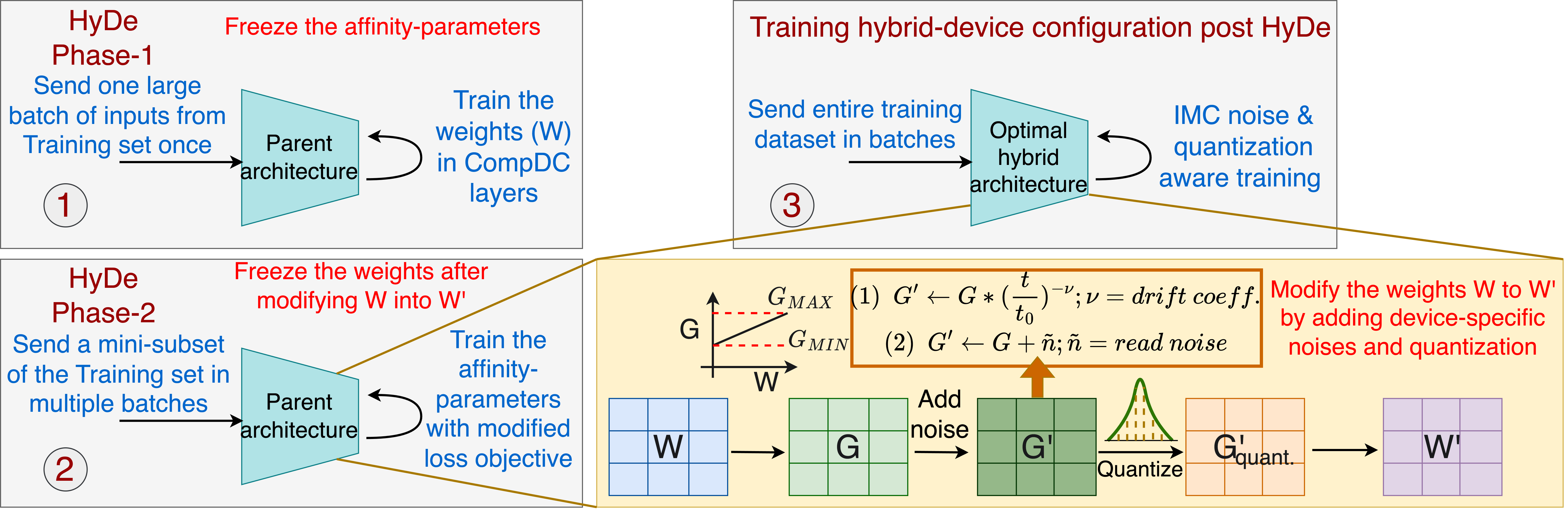}
   
    \caption{Representation of the HyDe framework followed by post-HyDe training of the optimal hybrid-device model. Phase-1 shows one-shot training of weights in the parent architecture using cross-entropy loss. Phase-2 shows inclusion of the device-specific characteristics on the weights of the parent architecture and then training the affinity-parameters with a modified loss objective function.}
    \label{hyde}

\end{figure*}

\subsection{Training Methodology of HyDe}
\label{sec:hyde_train}

Here, we discuss our HyDe framework which trains the device-specific affinity-parameters, thereby helping us deduce the optimal hybrid-device configuration for a given DNN model and task. We run the HyDe framework for 30 epochs. One epoch of training the parent architecture occurs in two phases (see Fig. \ref{hyde}) as follows:

\textbf{Phase-1: Training the weight parameters.} 
A batch of inputs with a large batch-size (1000 in this work) is randomly sampled from the Training dataset and forwarded through the parent architecture once. Thereafter, the weight parameters ($W$) for all the CompDC layers in the parent architecture are updated using backpropagation with standard cross-entropy loss ($\Lagr_{CE}$). Note, during Phase-1, we do not include any device-related parameters  and the affinity-parameters ($\alpha$) are unaffected in this phase.

\textbf{Phase-2: Training the affinity-parameters.}
We first incorporate the impact of the device-specific properties (listed in Table \ref{tab:dev_prop})-temporal drift, read noise and quantization on the weights ($W$) of the parent architecture in the manner shown in Fig. \ref{hyde}. We convert the weights into device conductances $G$ and then, incorporate device-noise in two steps. First, we include the impact of temporal conductance drift (if applicable) to the conductances using eq. \ref{eq:drift-noise}. Unless otherwise mentioned, the value of inference time $t$ is taken as 100s to compute drift noise. Second, we add read-noise to the conductances based on the values of $\sigma$ in Table \ref{tab:dev_prop} using eq. \ref{eq:read-noise}. Thus, we obtain noisy conductances $G'$ which are uniformly quantized based on the precisions in Table \ref{tab:dev_prop} to $G'_{quant.}$. Finally, $G'_{quant.}$ is transformed back to the weight-domain and we obtain the modified weights $W'$. Thereafter, the modified weights ($W'$) are kept frozen. We train the $\alpha$ parameters by feeding in inputs from a mini-subset of the training dataset (described in Section \ref{sec:expt}) using a modified loss objective function ($\Lagr_{mod}$) via backpropagation. Note, after training the $\alpha$ parameters, we restore the original $W$ in the parent architecture. 

We follow a differentiable approach \cite{cai2018proxylessnas} to optimize for IMC crossbar area and device-programming energy by regularizing our loss objective function. Say, for the $i^{th}$ CompDC layer, the total IMC area and Programming energy with the $j^{th}$ device (SRAM, PCM or FeFET) be $a_{ij}$ \& $e_{ij}$, respectively. Note that $a_{ij}$ \& $e_{ij}$ are computed using the individual device-specific areas and programming energies listed in Table \ref{tab:dev_prop}. Since, the constituent devices of a CompDC layer are associated with probability-coefficients ($p_{ij}$'s), the expected values of total IMC area ($E[IMC~area]_i$) and total programming energy ($E[Prog.~Energy]_i$) for $i^{th}$ CompDC layer are:\\ $E[IMC~area]_i = \sum_{j=1}^{3}{a_{ij}*p_{ij}}$, and $E[Prog.~Energy]_i  = \sum_{j=1}^{3}{e_{ij}*p_{ij}}$.

Our modified loss objective function ($\Lagr_{mod}$) is defined as:

\small
\begin{equation} \Lagr_{mod} = \Lagr_{CE} + \lambda_1*\sum_{\forall i}^{}{E[IMC~area]_i} +  \lambda_2*\sum_{\forall i}^{}{E[Prog.~Energy]_i}.
\label{eq:loss_mod}
\end{equation}

\normalsize

Here, $\lambda_1$ and $\lambda_2$ are hyperparameters that control the relative importance given to $E[IMC~area]$ \& $E[Prog.~Energy]$, respectively, with respect to $\Lagr_{CE}$.

\subsection{Sampling and Training the Optimal Hybrid-device Model}
\label{sec:hwt}

Finally, based on the trained affinity-parameters, we derive the optimal hybrid-device configuration from the parent architecture by choosing the device for a CompDC layer with the highest value of $\alpha$ or probability-coefficient. In other words, we modify eq. (\ref{eq:out-CompDC}) as:

\small
\begin{equation} m_{CompDC}  = \sum_{\forall j}^{}g_j * o_j,\quad g_j =
\begin{cases}
 1 ,   & p_j = max(p) \\
    0 , & otherwise.
\end{cases}
\label{eq:out-final}
\end{equation}
\normalsize

The optimal hybrid-device model's weight parameters (W) are then trained to convergence using backpropagation with layerwise device-awareness in the same manner as for Phase-2 search in HyDe. However this hardware-aware training is performed using standard cross-entropy loss based on the parameters in Table \ref{tab:dev_prop}, where we only inject device-specific read noise (without drift noise) into the weights and quantize the weights (programmed as 4-bit conductances) as well as the partial-sums arising from the ADCs. Please note that although in this work, we employ a standard IMC hardware-aware training method, we believe that including more matured hardware-aware training methods \cite{rasch2023hardware, krishnan2022exploring, wu2023bulk} will further improve the performance accuracies of the crossbar-mapped DNN architectures.

\section{Experiments and Results}
\label{sec:expt}

\begin{table*}[t]
\centering
\caption{Table showing overall results pertaining to the baselines and the HyDe-derived hybrid models for a VGG16 DNN. Note for the hybrid models, the layerwise configurations denote the choice of devices preferred by each conv2D layer in a sequential order from layer L1-L12.}
\label{tab:res_1}
\resizebox{\linewidth}{!}{%
\begin{tabular}{|c|c|c|c|c|c|c|c|}
\hline
\textbf{Task/Dataset} &
  \textbf{Model} &
  \textbf{\begin{tabular}[c]{@{}c@{}} HyDe-derived \\ Layerwise Configuration \\ (S = SRAM, P = PCM, F = FeFET)  \end{tabular}} &
  \textbf{\begin{tabular}[c]{@{}c@{}}\% Inference \\ accuracy \\ (t=0)\end{tabular}} &
  \textbf{\begin{tabular}[c]{@{}c@{}}IMC \\ area \\ ($mm^2$)\end{tabular}} &
  \textbf{\begin{tabular}[c]{@{}c@{}}Average \\ prog. energy \\ ($\mu J$)\end{tabular}} &
  \textbf{\begin{tabular}[c]{@{}c@{}}ADC \\ energy \\ ($\mu J$)\end{tabular}} &
  \textbf{$TOPS/mm^2$} \\ \hline
\textbf{CIFAR10} &
  Baseline SRAM &
  All SRAM &
  87.62 &
  15.6 &
  0.2 &
  19.7 &
  7.03 \\ \cline{2-6} \cline{8-8} 
 &
  Baseline PCM &
  All PCM &
  84.24 &
  0.13 &
  498 &
   &
  10.4 \\ \cline{2-6} \cline{8-8} 
 &
  Baseline FeFET &
  All FeFET &
  86.83 &
  0.19 &
  99.6 &
   &
  10.4 \\ \cline{2-8} 
 &
  \textbf{Hybrid-I} &
  F, S, P, P, F, P, F, F, P, F, F, F &
  87.11 &
  0.48 &
  199.2 &
  18.5 &
  12.8 \\ \cline{2-8} 
 &
  \textbf{Hybrid-II} &
  S, P, S, P, F, F, F, F, F, F, F, P &
  86.84 &
  0.773 &
  194.0 &
  18.27 &
  12.6 \\ \cline{2-8} 
 &
  \textbf{Hybrid-III} &
  S, P, S, P, P, F, F, F, F, F, F, P &
  87.36 &
  0.77 &
  152.1 &
  18.31 &
  12.5 \\ \hline
\textbf{TinyImagenet} &
  Baseline SRAM &
  All SRAM &
  52.62 &
  15.6 &
  0.2 &
  75.9 &
  2.37 \\ \cline{2-6} \cline{8-8} 
 &
  Baseline PCM &
  All PCM &
  47.32 &
  0.13 &
  498 &
   &
  2.81 \\ \cline{2-6} \cline{8-8} 
 &
  Baseline FeFET &
  All FeFET &
  50.88 &
  0.19 &
  99.6 &
   &
  2.82 \\ \cline{2-8} 
 &
  \textbf{Hybrid-I} &
  P, P, S, F, P, F, P, F, F, F, F, F &
  50.31 &
  0.483 &
  178.3 &
  56.0 &
  5.94 \\ \cline{2-8} 
 &
  \textbf{Hybrid-II} &
  F, S, F, P, P, F, F, F, F, F, F, F &
  50.64 &
  0.487 &
  136.3 &
  53.3 &
  6.23 \\ \cline{2-8} 
 &
  \textbf{Hybrid-III} &
  S, S, F, F, P, F, F, F, F, F, F, F &
  50.62 &
  0.785 &
  110.1 &
  52.4 &
  6.49 \\ \hline
\end{tabular}%
}

\end{table*}

\textbf{Experimental Setup:} We build our parent architecture using a VGG16 DNN (with 12 conv2D layers, namely L1-L12) and show experimental results on benchmark tasks or datasets- CIFAR10 and TinyImagenet. The layerwise network configuration of the VGG16 model is as follows:
\textbf{conv (3,64), conv (64,64), M, conv (64,128), conv (128,128), M, conv (128,256), conv (256,256), conv (256,256), conv (256,256), M, conv (256,512), conv (512,512), conv (512,512), conv (512,512), M, FC}.
Here, \textbf{M} stands for max-pooling layers (with stride of 2) and \textbf{FC} stands for the fully-connected classifier layer. The conv2D layers are represented as \textbf{conv (n\_i,n\_o)} where, n\_i \& n\_o denote the number of input \& output channels, respectively. The CIFAR10 dataset consists of RGB images (50,000 training and 10,000 testing) of size 32$\times$32 belonging to 10 classes. The TinyImagenet dataset is a more complex dataset with RGB images (100,000 training and 10,000 testing) of size 64$\times$64 belonging to 200 classes. 
Note, the mini-subset of training dataset (for Phase-2 training) is constructed by randomly sampling 10\% of the images from the training examples. 
We carry out the HyDe search for 30 epochs using Adam Optimizer with an initial learning rate of 0.06 for training the affinity-parameters. For the CIFAR10 (TinyImagenet) dataset, the optimal hybrid-device configuration models are trained to convergence for 40 (30) epochs with initial learning rate = $1e-3$ ($1e-4$) using Adam Optimizer. 

For each dataset, we generate three optimal hybrid-device architectures using HyDe- \textbf{Hybrid-I}, \textbf{Hybrid-II} \& \textbf{Hybrid-III}. This is done by either adjusting the relative importances of $\lambda_1$ and $\lambda_2$ hyperparameters against $\Lagr_{CE}$ in eq. \ref{eq:loss_mod} (Hybrid-I \& II) or by increasing Inference time ($t$) in Phase-2 search for better retention capabilities (Hybrid-III). Hybrid-I denotes the model optimized with higher importance to overall IMC area. Hybrid-II denotes the model optimized with higher importance to overall device-programming energy. Hybrid-III is generated by increasing $t$ to 1000s from 100s. Note, our baselines for comparison are full homogeneous SRAM, PCM and FeFET-based architectures. All the baselines and the hybrid-device models are trained to convergence with hardware-aware training (see Section \ref{sec:hwt}), where we consider 4-bit ADCs for the CIFAR10 dataset and 6-bit ADCs for the TinyImagenet dataset to quantize the partial-sums. We build a wrapper on top of the Neurosim tool \cite{chen2018neurosim} to include the impact of layerwise device-heterogeneity and compute the hardware metrics (energy, area and $TOPS/mm^2$) corresponding to the baselines and the hybrid architectures. Note, Neurosim is a Python-based hardware-evaluation tool that performs a holistic energy-latency-area evaluation of analog crossbar-based DNN accelerators implemented on a tiled architecture similar to Fig. \ref{overall_fig}(b). We calibrate our tool for the device properties in Table \ref{tab:dev_prop} with a crossbar size of \textbf{128$\times$128} and \textbf{64 PEs per tile} in 32 nm CMOS technology node, unless stated otherwise.

Please note that although we derive hybrid-device architectures for the VGG16 topology using HyDe, we can use the HyDe framework for any fixed DNN model (such as ResNet, DenseNet, \textit{etc.}). It would only require us to re-architect the parent architecture according to the new DNN topology and then run HyDe on the same to get the optimal hybrid-device architecture.

\subsection{Overall Results}

\begin{figure}[t]
    \centering
    \includegraphics[width=\linewidth]{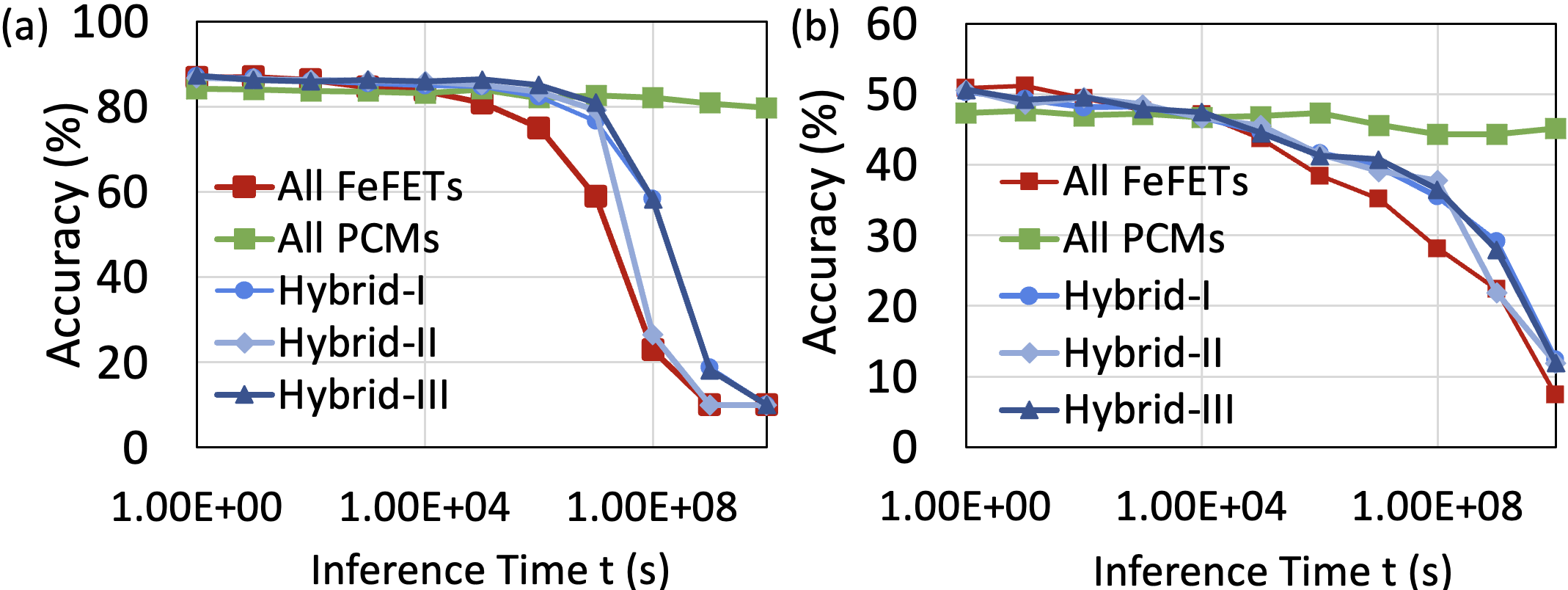}
   
    \caption{Plots for VGG16 Inference accuracy versus time for the baselines and the hybrid-device models indicating the retention of each architecture using- (a) CIFAR10, (b) TinyImagenet datasets.}
    \label{retention}

\end{figure}

Table \ref{tab:res_1} shows that although the `All SRAM' baseline has the best performance in terms of inference accuracy and low programming energy, it has a huge IMC area overhead and low $TOPS/mm^2$. While the IMC area overhead gets reduced in the `All PCM' baseline by $\sim120\times$, it suffers from high accuracy degradation ($\sim3.4-5.3\%$) due to the impact of read noise and higher device-programming energy ($>2000\times$). The `All FeFET' baseline, having lower read noise, achieves better test accuracy at a lower cost of programming and with a slightly higher area ($\sim1.5\times$) overhead than the `All PCM' baseline. However, the `All FeFET' baseline has the poorest retention owing to the highest temporal drift as shown in Fig. \ref{retention}. In Fig. \ref{retention}, the `All PCM' baseline has the best retention ($>10~years$) but has a lower initial test accuracy at $t=0$ due to the impact of higher read noise. Our HyDe-derived hybrid-device models achieve classification accuracy close to the `All SRAM' baselines, while optimizing for area-efficiency ($\sim20-32\times$ more IMC area-efficient) and programming energies by reducing the average energy to $110.1\mu J$. Specifically, the latter conv2D layers prefer FeFETs to reduce the impact of read noise, while maintaining optimal area-efficiency. Our hybrid models also have optimal retention capabilities (Hybrid-III being the best with reasonably low device-programming energies) as shown in Fig. \ref{retention}. This is done by preferring SRAM/PCM devices in the initial layers.
The discussion on increased $TOPS/mm^2$ and reduced inference energies of the hybrid models is presented in Section \ref{sec:adc}.

\begin{figure}[t]
    \centering
    \includegraphics[width=\linewidth]{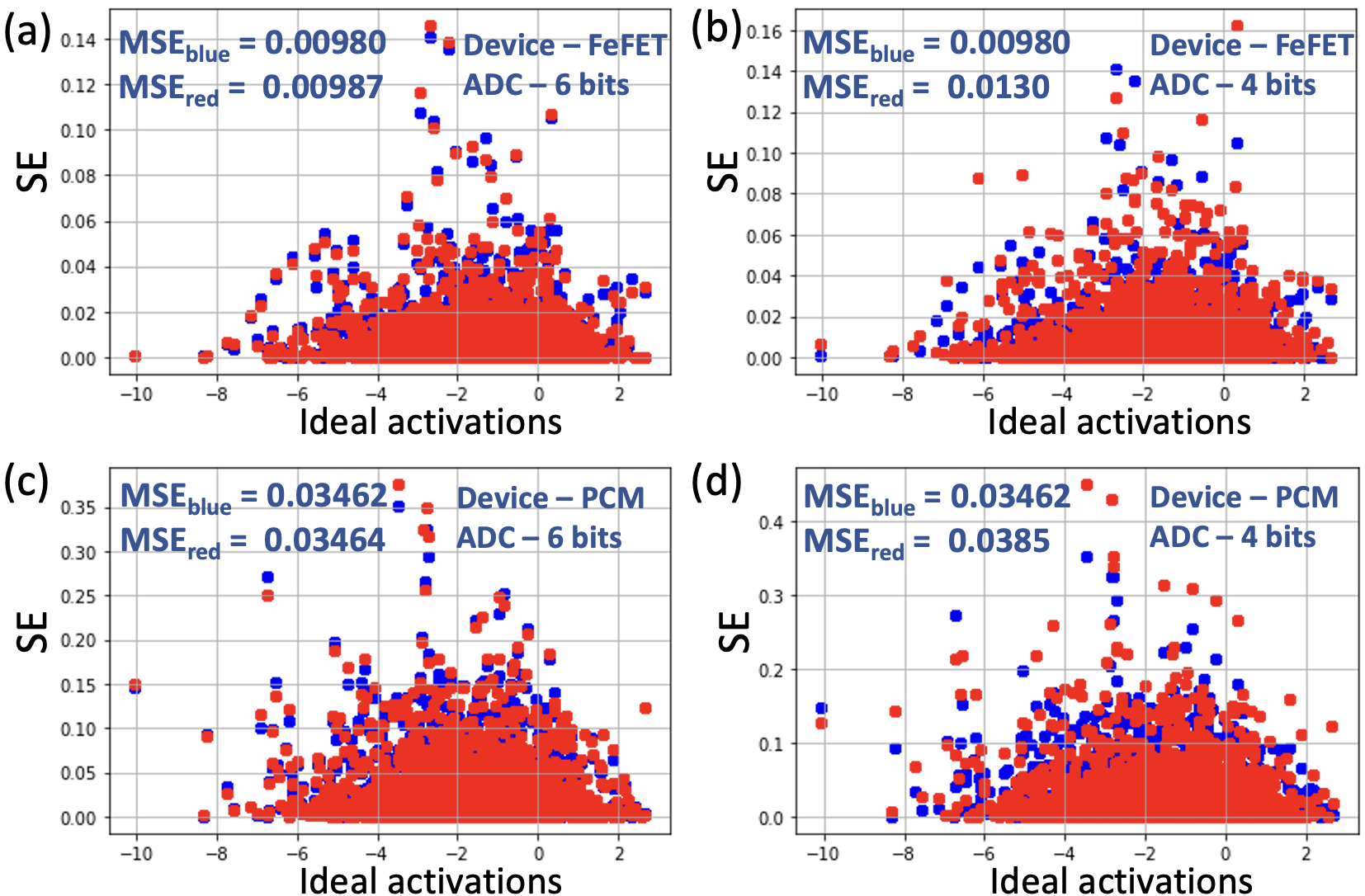}
   
    \caption{Plot of SEs versus ideal activations of L5 layer in a pre-trained VGG16 DNN for- (a) weights on FeFET crossbars and ADC precision of 6 bits, (b) weights on FeFET crossbars and ADC precision of 4 bits, (c) weights on PCM crossbars and ADC precision of 6 bits, (d) weights on PCM crossbars and ADC precision of 4 bits. Here, blue points denote SEs for noisy output activations without ADC quantization, while red points denote SEs post ADC quantization.}
    \label{noise_ADC}
     
\end{figure}

\begin{table*}[ht]
\centering
\caption{Table showing the transferability of HyDe-derived hybrid models across datasets (transfer from CIFAR10 to TinyImagenet).}
\label{tab:trans_res}
\resizebox{\linewidth}{!}{%
\begin{tabular}{|c|c|c|c|c|}
\hline
\textbf{Model} &
  \textbf{\%Inference accuracy (t=0)} &
  \textbf{Average prog. energy ($\mu J$)} &
  \textbf{Inference energy ($\mu J$)} &
  \textbf{$TOPS/mm^2$} \\ \hline
Hybrid-III &
  50.62 &
  110.1 &
  125.0 &
  6.49 \\ \hline
\begin{tabular}[c]{@{}c@{}}Hybrid-III \\ (T/F from CIFAR10)\end{tabular} &
  50.84 &
  152.1 &
  128.0 &
  6.04 \\ \hline
\end{tabular}%
}
\vspace{-4mm}
\end{table*}

\subsection{Impact on ADC Precisions and Overall $TOPS/mm^2$}
\label{sec:adc}

One interesting consequence of using layerwise hybrid devices is that the layerwise ADC precisions can be varied depending upon the choice of the device. Let us understand this from the analysis shown in Fig. \ref{noise_ADC}. We know that FeFET is less impacted by read noise than PCM due to lower $\sigma$ and higher ON/OFF ratio (see Table \ref{tab:dev_prop}). For Fig. \ref{noise_ADC}(a-b),  we take a pre-trained VGG16 model and map layer L5 using FeFET devices. We plot the squared-errors (SEs) between the noisy and ideal output activations against the ideal output activations (blue points). Thereafter, we quantize the noisy output activations using 6-bit (see Fig. \ref{noise_ADC}(a)) and 4-bit (see Fig. \ref{noise_ADC}(b)) ADCs, respectively, and plot their SEs against the ideal output activations (red points). In Fig. \ref{noise_ADC}(c-d), we repeat the same procedure but using PCM devices. For all the plots, we report the mean-squared-errors (MSEs) for the blue and the red points as $MSE_{blue}$ and $MSE_{red}$, respectively. Note, the activations input to the L5 layer are the same for all the plots in Fig. \ref{noise_ADC} and generated from a random uniform distribution for a fair comparison. Clearly, $MSE_{blue}$ for PCM-generated activations is higher than that of the FeFET-generated activations due to higher impact of PCM read noise. Quantizing the noisy activations using 4-bit ADCs deteriorates the MSEs further in case of PCM devices to 0.0385, while for FeFET devices it is still lower at 0.0130. This shows that PCM devices that are more susceptible to read noise should operate in conjunction with a higher ADC precision than FeFETs. 

Based on the above observation, we set the layerwise ADC precision in the HyDe-derived hybrid IMC models depending upon the type of device chosen. PCM-based layers require ADC precision higher than FeFETs, while the SRAM-based layers can have the lowest ADC precision (as they are least impacted by stochastic read noise). With lower ADC precisions, we can significantly reduce the overall chip area as ADC accounts for a major share of the overall chip area as shown in Fig.\ref{intro_fig}(a). This results in increased $TOPS/mm^2$ for the hybrid-device models (see Table \ref{tab:res_1}). In addition, the overall inference energy is also lowered in case of the hybrid models, owing to reduction in the ADC energy. Note, the homogeneous baseline models are based on uniform ADC precisions across all layers. 

\begin{figure}[t]
    \centering
    
    \includegraphics[width=\linewidth]{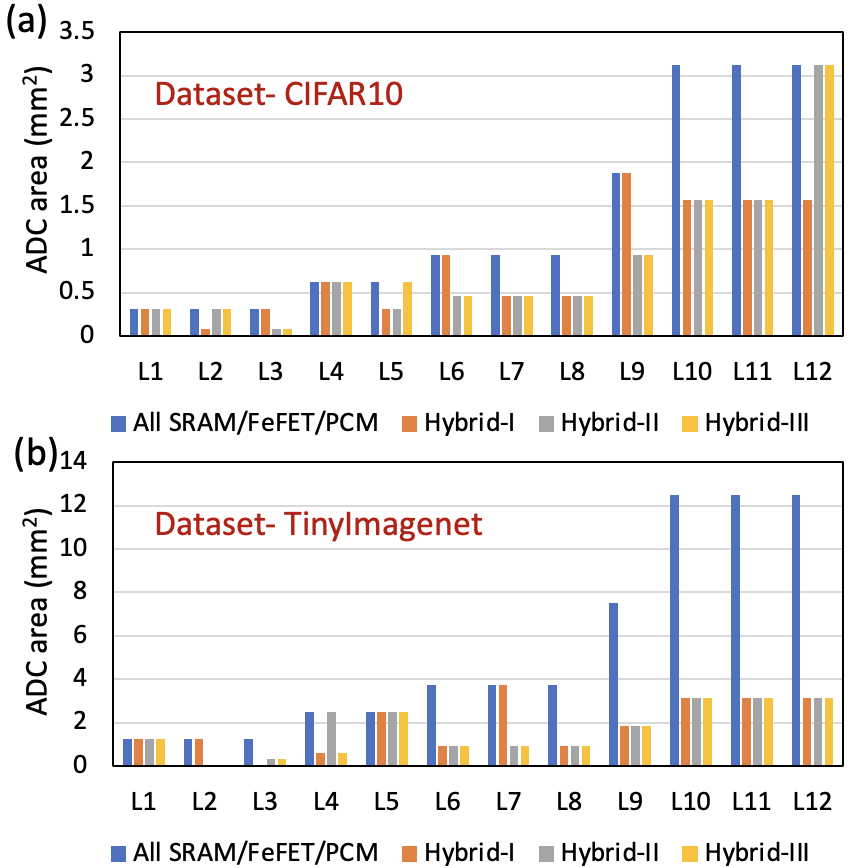}%

    \caption{Plots for layerwise (L1-L12) ADC area for the baselines and the hybrid-device architectures using- (a) CIFAR10, (b) TinyImagenet datasets.}
    \label{adc_cifar10_tiny}
    \vspace{-5mm}
\end{figure}

For the CIFAR10 (TinyImagenet) dataset, the minimum ADC precision to maintain the classification performance was found to be 4 bits (6 bits). Hence, all the baseline homogeneous models have 4-bit (6-bit) ADC precisions. For the hybrid models, barring the L1 layer, all SRAM-based layers have 2-bit (2-bit) ADCs, FeFET-based layers have 3-bit (4-bit) ADCs and PCM-based layers have 4-bit (6-bit) ADCs. The first layer irrespective of the nature of device requires 4-bit (6-bit) ADCs to maintain performance. In Fig. \ref{adc_cifar10_tiny}, as the latter conv2D layers (L5-L12) generally prefer FeFETs, the reduction in ADC precision brings huge area savings for the hybrid models against the baselines ($\sim4-16\times$). This boosts the $TOPS/mm^2$ of the hybrid-device models by $\sim1.82\times$ for CIFAR10 and $\sim2.74\times$ for TinyImagenet tasks with respect to the `All SRAM' baseline. We also find a $\sim8\%$ and $\sim22-26\%$ reduction in the overall inference energy due to the layerwise heterogeneity in ADC precisions for CIFAR10 and TinyImagenet datasets, respectively.

\subsection{Transferability Across Similar Tasks}

\begin{figure}[t]
    \centering
    \includegraphics[width=.65\linewidth]{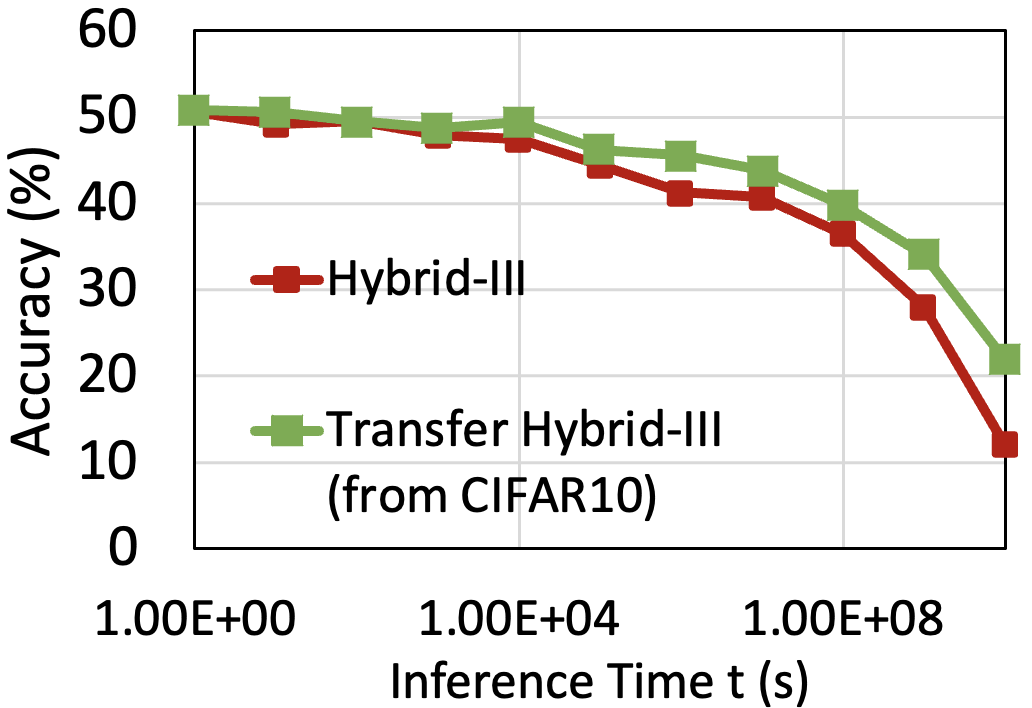}
   
    \caption{Plot showing the retention of Hybrid-III model for TinyImagenet versus the Hybrid-III model derived using CIFAR10 and transferred to TinyImagenet dataset.}
    \label{ret_trans}
     
\end{figure}

Table \ref{tab:trans_res} and Fig. \ref{ret_trans} show that a CIFAR10-generated hybrid-device configuration transfers well to a more complex TinyImagenet task. This \textbf{eliminates chip re-fabrication costs} for deploying a hybrid IMC architecture pertaining to a given DNN topology for a different task.  
The VGG16 DNN trained on Hybrid-III configuration transferred from CIFAR10 is \textbf{iso-accurate} with the Hybrid-III model for TinyImagenet and shows \textbf{better retention}, with marginally higher inference energy and lower $TOPS/mm^2$. Also, since TinyImagenet is a larger dataset (100,000 training images) with larger image size (64$\times$64), the total training time per epoch for HyDe search is $>8\times$ higher than its CIFAR10 counterpart. With transferability, we need not re-run HyDe for the TinyImagenet dataset, thereby \textbf{saving on the overall training time}.

\subsection{Impact of Resistive Non-idealities \& Crossbar Size}

\begin{wrapfigure}{l}{0.25\textwidth}
\includegraphics[width=0.25\textwidth]{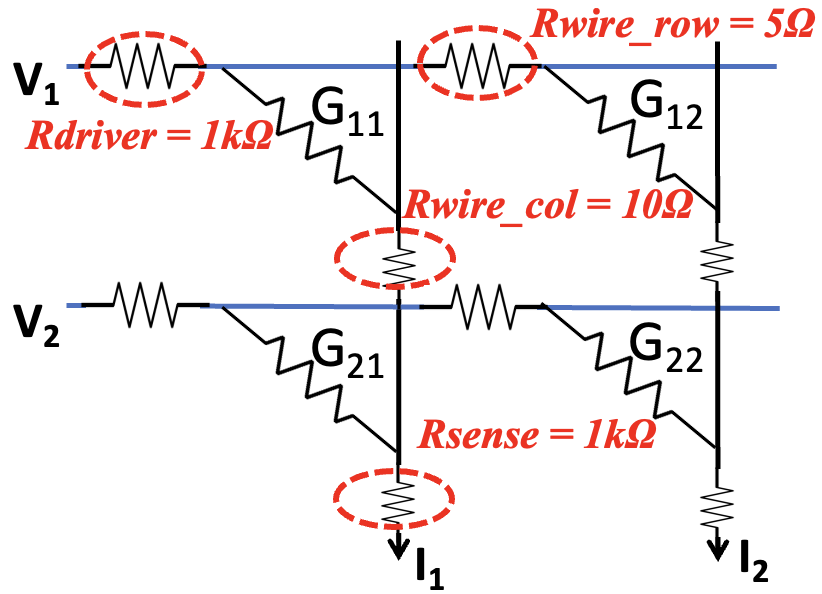}
  
\caption{A 2$\times$2 IMC crossbar with resistive non-idealities annotated in red. }
\label{xbar_IRdrop}
\vspace{-2mm}
\end{wrapfigure}

Here, we include the resisitive parasitic non-idealities (annotated in Fig. \ref{xbar_IRdrop} with values) in crossbars during inference \cite{bhattacharjee2021neat, bhattacharjee2021efficiency}. The impact of the resistive non-idealities on the DNN weights is simulated using the RxNN framework \cite{jain2020rxnn}. Prior works \cite{bhattacharjee2021neat, bhattacharjee2020switchx} have shown that larger crossbars are invariably more prone to the impact of resistive non-idealities, thereby degrading the accuracy of the mapped DNNs. Furthermore, crossbars having IMC devices with higher $R_{ON}$ help reduce the impact of such non-idealities \cite{bhattacharjee2021neat, moitra2023spikesim}.

\begin{figure}[b]
    \centering
    \includegraphics[width=\linewidth]{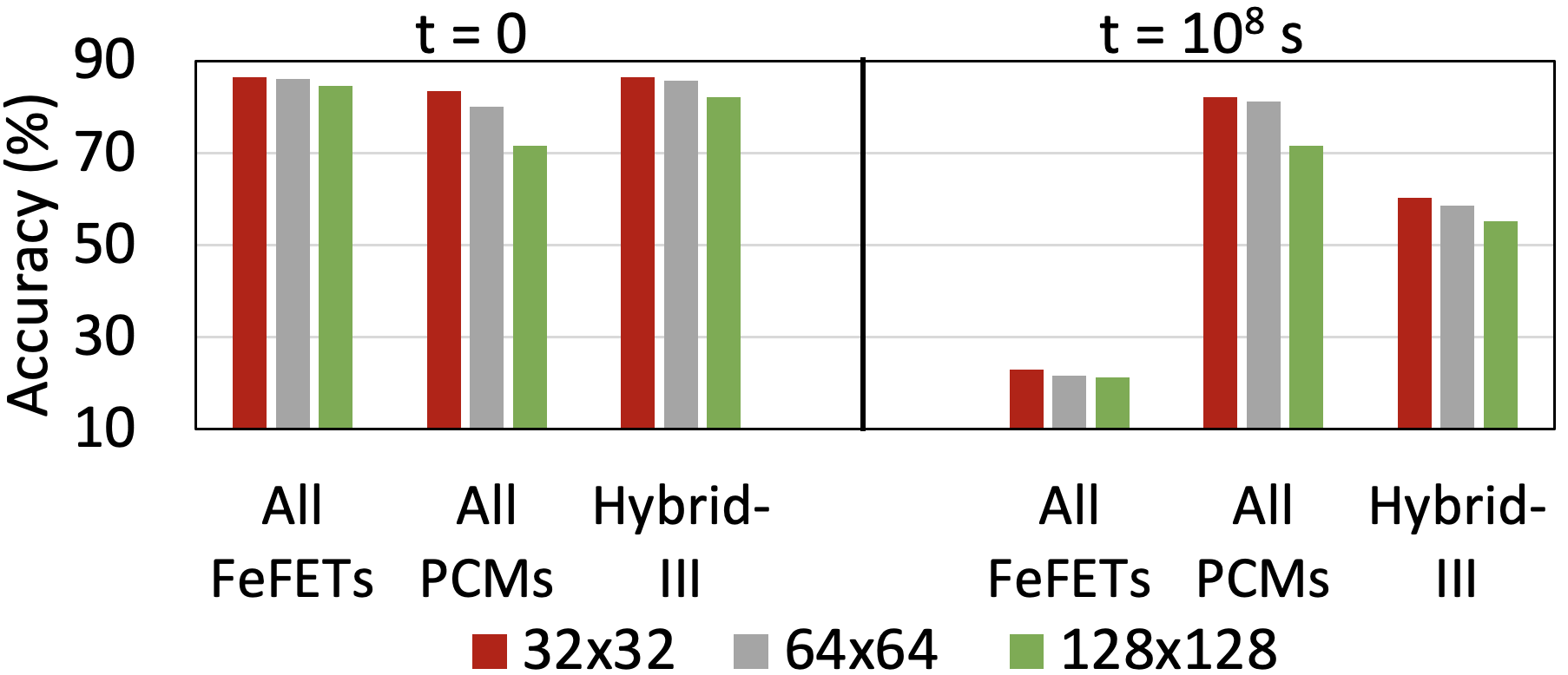}
   
    \caption{Plot showing the impact of resistive non-idealities across crossbar sizes on the CIFAR-10 test accuracy of homogeneous baselines and hybrid-device models at $t=0$ \& $t=10^8s$.}
    \label{ret_IRdrop}
     
\end{figure}

\begin{figure*}[t]
    \centering
    \subfloat[]{
    \includegraphics[width=.35\linewidth]{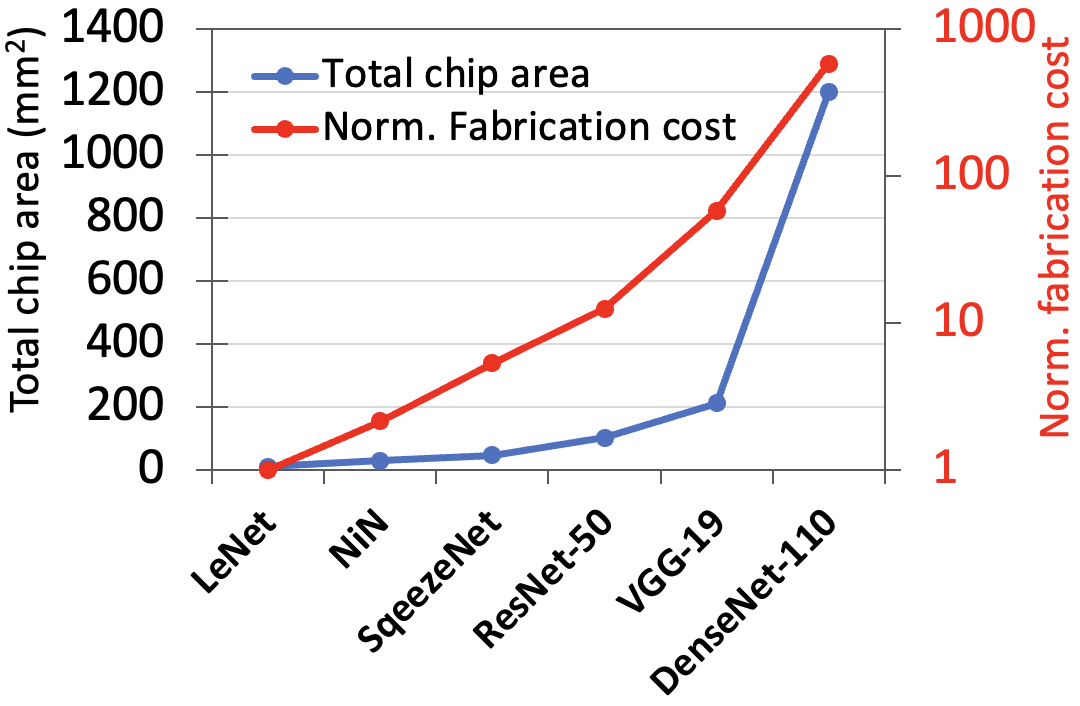}%
    } 
    \subfloat[]{
    \includegraphics[width=.65\linewidth]{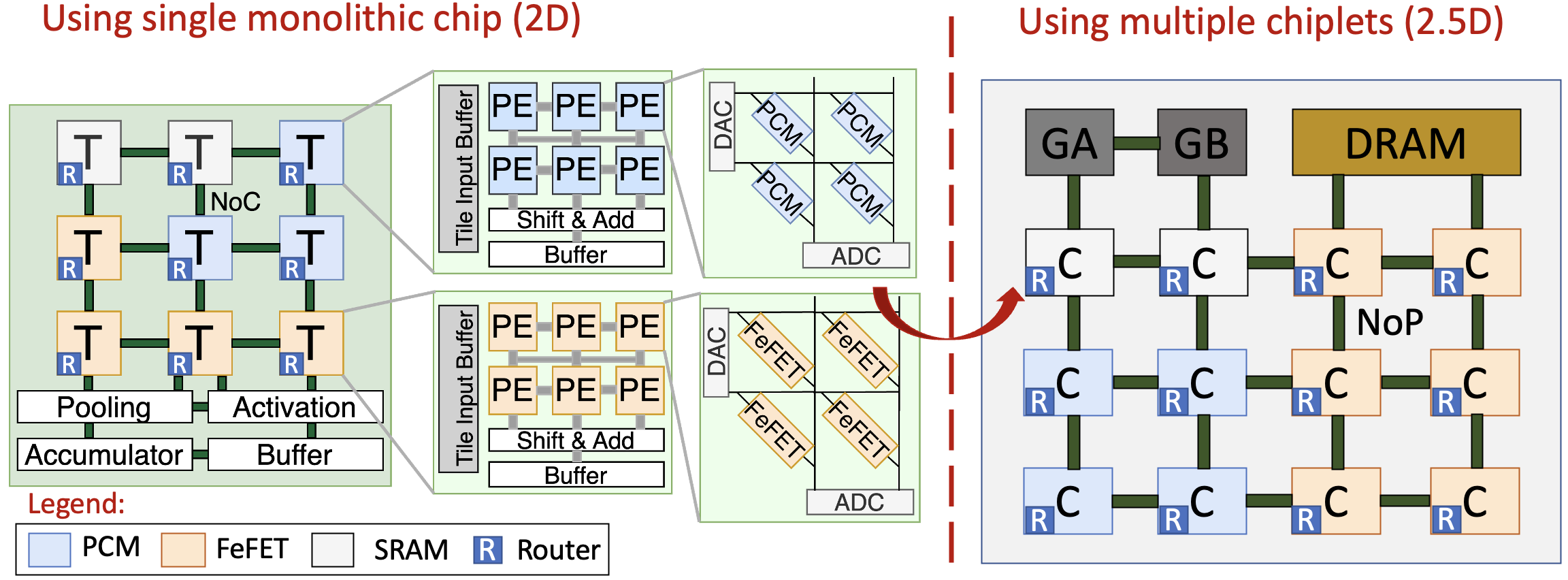}
    }
    \caption{(a) Total monolithic chip area and fabrication cost (normalized by the fabrication cost of an LeNet model) for an `All PCM' IMC architecture across DNN models of increasing size (data adapted from \cite{krishnan2021siam}). (b) Proposed implementation of the HyDe-derived multi-device IMC models in the 2.5D design space using an ensemble of device-specific IMC chiplets (C). }
    \label{2-2.5D}

\end{figure*}

\begin{table*}[t]
\centering
\caption{Table comparing recent works based on IMC crossbar-based implementations of DNNs in the 2.5D/3D design space against our work. Here, N.A. stands for Not Applicable.}
\label{tab:comparison}
\resizebox{\linewidth}{!}{%
\begin{tabular}{|c|c|c|c|c|c|c|}
\hline
\textbf{Work} &
  \textbf{\begin{tabular}[c]{@{}c@{}}Integration \\ mode\end{tabular}} &
  \textbf{Technology} &
  \textbf{\begin{tabular}[c]{@{}c@{}}Heterogeneity \\ in architecture\end{tabular}} &
  \textbf{\begin{tabular}[c]{@{}c@{}}Optimizing \\ IMC energy \& \\ area-efiiciencies\end{tabular}} &
  \textbf{\begin{tabular}[c]{@{}c@{}@{}}Optimizing \\ DNN accuracy \& \\ retention under \\ IMC noise\end{tabular}} &
  \textbf{\begin{tabular}[c]{@{}c@{}}Optimizing \\ NoC/NoP \\ energy-efficiencies\end{tabular}} \\ \hline
{\cite{krishnan2021siam}} &
  2.5D &
  RRAM+CMOS &
  \begin{tabular}[c]{@{}c@{}}N.A. \\ (Homogeneous \\ chiplet-sizes)\end{tabular} &
  \textbf{\checkmark} &
  N.A. &
  \textbf{\checkmark} \\ \hline
{\cite{krishnan2021system, sharma2023achieving}} &
  2.5D &
  \begin{tabular}[c]{@{}c@{}}RRAM+CMOS \\ or FeFET+CMOS \\ or PCM+CMOS \end{tabular} &
  \begin{tabular}[c]{@{}c@{}}N.A. \\ (Homogeneous \\ chiplet-design \& sizes)\end{tabular} &
  N.A. &
  N.A. &
  \textbf{\checkmark} \\ \hline
{\cite{wang2022ai}} &
  2.5D &
  RRAM+CMOS &
  \begin{tabular}[c]{@{}c@{}}Uses layer-specific chiplets of \\ varying sizes \\ (Big-little chiplets)\end{tabular} &
  \textbf{\checkmark} &
  N.A. &
  \textbf{\checkmark} \\ \hline
{\cite{murali2020heterogeneous}} &
  3D &
  RRAM+CMOS &
  \begin{tabular}[c]{@{}c@{}}Uses two-tiered \\ 3D monolithic chips \\ at different technology nodes\end{tabular} &
  \textbf{\checkmark} &
  N.A. &
  N.A. \\ \hline
\textbf{Our work} &
  \textbf{2.5D} &
  \textbf{\begin{tabular}[c]{@{}c@{}@{}}CMOS (incl. SRAM IMC) \\ + FeFET \\ + PCM\end{tabular}} &
  \textbf{\begin{tabular}[c]{@{}c@{}}Heterogeneous chiplets \\ using IMC devices \& \\ associated ADC precisions\end{tabular}} &
  \textbf{\textbf{\checkmark}} &
  \textbf{\textbf{\checkmark}} &
  \textbf{N.A.} \\ \hline
\end{tabular}%
}

\vspace{-4mm}
\end{table*}

Fig. \ref{ret_IRdrop} shows the inference accuracy (CIFAR10) of the baselines and hybrid models across different crossbar sizes. At $t=0$, the `All FeFET' baseline with the highest $R_{ON}$ suffers the least accuracy loss ($\sim2\%$ loss) , while the `All PCM' baseline (lower $R_{ON}$) has the least inference accuracy and incurs the highest degradation at higher crossbar size of 128$\times$128 ($\sim12\%$ loss). Our Hybrid-III model achieves iso-accuracy with the `All FeFET' baseline for 32$\times$32 \& 64$\times$64 crossbars and reduces the accuracy loss to $<5\%$ at higher crossbar size of 128$\times$128. At $t=10^8s$, the `All FeFET' baseline with poor retention, achieves close to $20\%$ accuracy. The `All PCM' baseline has the best accuracy due to the higher retention but suffers $\sim12\%$ accuracy loss at 128$\times$128 crossbars. Our Hybrid-III model maintains $\sim60\%$ test accuracy with a degradation of $\sim5\%$ at 128$\times$128 crossbars.

\section{Proposed Implementation of HyDe-derived Architectures in the 2.5D Design Space}

So far we have shown results upon implementing the hybird-device IMC architectures on a single monolithic chip. However, the fabrication complexity and design effort for a such a chip increases as multiple processes are involved in the fabrication of multi-device IMC architectures. In addition, larger and branched DNN topologies with higher on-chip memory requirements can incur huge on-chip area, and thus fabrication costs. From Fig. \ref{2-2.5D}(a), we find that the fabrication costs (plotted in logarithmic scale) increases exponentially with the chip area, thereby making monolithic chip-design less cost-efficient, especially for larger DNNs, if all the DNN parameters are to be stored on a single chip \cite{krishnan2021siam}. Larger chips with hybrid IMC device integration will suffer from lower yield and higher defects across the wafer owing to multiple design processes involved, which will exacerbate the fabrication costs \cite{stow2016cost}. 

To circumvent the above challenges with monolithic integration, 2.5D integration using chiplet-based IMC architectures can come handy, leading to $\sim10-50\%$ reduction in the fabrication costs \cite{krishnan2021siam}. While 2.5D von-Neumann inference accelerators have been proposed using chiplets containing systolic-arrays of digital dot-product engines \cite{shao2019simba, liu2021256gb}, there lies great promise in the domain of 2.5D analog crossbar-based IMC accelerators which have shown $>24\times$ higher energy-efficiency \cite{krishnan2021siam} against their von-Neumann counterpart \cite{shao2019simba} in accelerating identical DNN workloads. To this end, in this work, we propose fabricating IMC device-specific chiplets, where each chiplet (C) consists of arrays of crossbars  of a particular device-type (SRAM, PCM or FeFET) and associated peripheral circuits. As shown in Fig. \ref{2-2.5D}(b), a large DNN model can be  deployed on multiple smaller chiplets, wherein a single DNN layer is built upon one or more independently fabricated chiplets of a specific device-type determined by the HyDe framework. All the chiplets along with the global accumulator (GA), global buffer (GB) and DRAM are integrated into a Package and interconnected using an Network-on-Package (NoP) that uses package-level signalling to connect different chiplets \cite{turner2018ground, lin20207}. Using an ensemble of small-sized device-specific chiplets helps improve the design effort and yield, reduces defect ratio and the fabrication cost in supporting multiple process technologies, thereby making our proposal for heterogeneous device-integration in the hardware implementation of deep learning workloads more pragmatic. Additionally, the 2.5D heterogeneous chiplet architecture is scalable to larger and complex DNN topologies as their constituent layers can be built by integrating chiplets of a particular design process \cite{krishnan2021siam}.  

Table \ref{tab:comparison} presents a qualitative comparison of our work in relation to previously proposed IMC implementations of deep learning workloads in the 2.5D/3D roadmap, highlighting our key contributions. 

\vspace{-4mm}
\section{Conclusion}

We propose HyDe framework that searches the layerwise affinity of a DNN for different IMC devices (PCM/FeFET/SRAM) to maintain high inference accuracy and maximize hardware area and energy-efficiencies. We find our HyDe-derived hybrid models to achieve upto $2.30-2.74\times$ higher $TOPS/mm^2$  at $22-26\%$ higher energy-efficiencies than baseline homogeneous architectures. With the emergence of the 2.5D chiplet technology to facilitate heterogeneous integration on hardware, we believe our work highlighting the benefits of hybrid IMC design using multiple devices is encouraging and timely.

\section*{Acknowledgement}

This work was supported in part by CoCoSys, a JUMP2.0 center sponsored by DARPA and SRC, Google Research Scholar Award, the NSF CAREER Award, TII (Abu Dhabi), the DARPA AI Exploration (AIE) program, and the DoE MMICC center SEA-CROGS (Award \#DE-SC0023198).

\bibliographystyle{IEEEtran}
\bibliography{reference}

\end{document}